# On the radiative heat exchange between spherical particles at small distances

G.V.Dedkov[1] and A.A.Kyasov

Kabardino –Balkarian State University, Nalchik, 360004, Russian Federation

**Abstract**
Radiative heat exchange of spherical particles is recalculated in the framework of fluctuation electrodynamics using the dipole approximation. We show that correct numerical coefficient in the resulting integral expression equals $4/\pi$, in contrast with that obtained by other authors.

A problem of radiative heat transfer between spherical dipole particles with different temperatures embedded in vacuum has been considered by several authors in the dipole approximation of fluctuation electrodynamics [1-6]. The restrictions imposed by this approach are:

$$R, R_2 << R; \quad \max(R_1, R_2) << \min(\lambda_{T1}, \lambda_{T2}) \tag{1}$$

where $R_{1,2}$ are the particles radii, $\lambda_{T1}$ and $\lambda_{T2}$ are the characteristic wave lengths of thermal radiation, and $R$ is the distance between the centres of the particles. In the case of particles modelled by fluctuating electric dipoles, the heating (cooling) rate of the first particle is given by

$$Q_{1\to 2}/dt = \chi \cdot \frac{\hbar}{R^6} \int_0^\infty d\omega \, \omega \alpha_1''(\omega) \alpha_2''(\omega) \left[3 + (\omega R/c)^2 + (\omega R/c)^4\right] \cdot$$
$$\cdot \left[\frac{1}{\exp(\hbar\omega/k_B T_2) - 1} - \frac{1}{\exp(\hbar\omega/k_B T_1) - 1}\right] \tag{2}$$

where $\alpha_{1,2}''(\omega)$ are the imaginary parts of the dipole polarizabilities, $\hbar, k_B$ are the Planck's and Boltzmann's constants, $\chi$ is numerical coefficient on the order of unit. An expression for $dQ_{2\to 1}/dt$ is given by the same eq.(2) replacing $T_1 \leftrightarrow T_2$. Note that if thermal radiation emitted to or absorbed from the vacuum background is also taken into account, than the corresponding contribution must be added in the right hand side of eq.(2) [5,6].

---

[1] Corresponding author e-mail: gv_dedkov@mail.ru

The most striking feature of eq.(2), as obtained by different authors, is a very large difference between the values of $\chi$, whereas the involved integral expressions prove to be the same. So, in Ref. [1] the authors have got $\chi = 1/2\pi^2$, while in a more recent paper of these authors one finds $128/\pi$ [2]; $1/4\pi^3$ in Ref.[3] and $1/8\pi^3$ in Ref.[4]. In our papers [5,6], on the other hand, we have got $\chi = 4/\pi$. As a matter of fact, the lack of concord in the value of $\chi$ may lead to a possible change of the heating rate by $10^4$ times !

This short note aims at solving this puzzle. It is worthwhile noting that correct calculation of the radiation mediated heat transfer (even in the dipole approximation) is of great importance in view of recent experimental observations [7-9].

In order to simplify the derivation as more as possible, we calculate the rate of radiative heat transfer between two point electric dipoles using the non retarded approximation. This allows to get correct expression for the leading term $dQ_{1 \to 2}/dt \propto R^{-6}$, and finally, for coefficient $\chi$. Our starting expression for $dQ_{1 \to 2}/dt$ completely agrees with [1-4] and can be cast in the form

$$\dot{Q}_{12} = \langle \dot{\mathbf{d}}_1^{in}(t) \mathbf{E}_2^{sp}(\mathbf{r}_1,t) \rangle - \langle \dot{\mathbf{d}}_2^{in}(t) \mathbf{E}_1^{sp}(\mathbf{r}_2,t) \rangle \tag{3}$$

where $\mathbf{d}_{1,2}(t)$ denotes dipole moments of the particles and $\mathbf{E}_{1,2}(\mathbf{r}_{2,1},t)$ denotes the fluctuation electromagnetic fields generated by each particle at the location point of another one. Superscripts "*in*", "*sp*" denote induced and spontaneous components, angular brackets –total quantum and statistical averaging. The first addend (3) corresponds to the heat produced in the volume of the first particle caused by spontaneous field of the second particle, while the second addend (3) denotes the heat produced in the volume of the second particle caused by spontaneous field of the first one. Making use Fourier representation for $\mathbf{d}_{1,2}(t)$ and $\mathbf{E}_{1,2}(\mathbf{r}_{2,1},t)$, we write

$$\dot{\mathbf{d}}_1^{in}(\omega) = -i\alpha_1(\omega) \mathbf{E}_2^{sp}(\mathbf{r}_1,\omega), \; \dot{\mathbf{d}}_2^{in}(\omega) = -i\omega\alpha_2(\omega) \mathbf{E}_1^{sp}(\mathbf{r}_2,\omega) \tag{4}$$

Moreover, in the non retarded approximation,

$$E_{1,i}^{sp}(\mathbf{r}_2,\omega) = T_{ij}(\mathbf{R})d_{1,j}^{sp}(\omega), \; E_{2,i}^{sp}(\mathbf{r}_1,\omega) = T_{ij}(\mathbf{R})d_{2,j}^{sp}(\omega) \tag{5}$$

$$T_{ij}(\mathbf{R}) = \frac{3n_i n_j - \delta_{ij}}{R^3}, \; \mathbf{n} = \mathbf{R}/R \tag{6}$$

Inserting (4)-(6) into $\langle \dot{\mathbf{d}}_1^{in}(t) \mathbf{E}_2^{sp}(\mathbf{r}_1,t) \rangle$ yields

$$dQ_{1\to 2}^{(1)}/dt = T_{ij}(\mathbf{R})T_{ik}(\mathbf{R})\int_{-\infty}^{\infty}\frac{d\omega}{2\pi}\int_{-\infty}^{\infty}\frac{d\omega'}{2\pi}(-i\omega)\exp[-i(\omega+\omega')t]\cdot\alpha_1(\omega)\cdot$$
$$\cdot\left\langle d_{2,j}^{sp}(\omega)d_{2,k}^{sp}(\omega')\right\rangle \qquad (7)$$

Using the correlator of fluctuating dipole moment [10]

$$\left\langle d_{2,j}^{sp}(\omega)d_{2,k}^{sp}(\omega)\right\rangle = 2\pi\delta_{jk}\delta(\omega+\omega')\hbar\alpha_1''(\omega)\coth(\omega\hbar/2k_B T_2), \qquad (8)$$

and relation $\sum_{i,j}(3n_i n_j - \delta_{ij})^2 = 6$, we finally get

$$dQ_{1\to 2}^{(1)}/dt = \frac{12}{\pi}\frac{1}{R^6}\int_0^{\infty}d\omega\,\omega\,\alpha_1''(\omega)\alpha_2''(\omega)\frac{1}{\exp(\hbar\omega/k_B T_2)-1} \qquad (9)$$

The term $\left\langle \dot{\mathbf{d}}_2^{in}(t)\mathbf{E}_1^{sp}(\mathbf{r}_2,t)\right\rangle$ is calculated quite analogously, and the result is given by eq.(9) after an obvious replacement $T_2 \to T_1$. Comparing eqs.(1) and (9) we see that $\chi = 4/\pi$, just as has been previously obtained in [5,6], using retarded fluctuation electrodynamics.